# Grover's search with an oracle distinguishing between solutions

*Hristo Tonchev[1], Rosen Bahtev[1]*

[1] *Institute for Nuclear Research and Nuclear Energy, Bulgarian Academy of Sciences, 72 Tzarigradsko Chaussée, 1784 Sofia, Bulgaria*
Emails: htonchev@ inrne.bas.bg

**Abstract:** *Here we suggest a modification of Grover's algorithm, based on a multiphase oracle which marks each solution with a different phase when there is more than one solution. Such a modification can be used to maintain a high probability of finding a solution for a number of iterations equal to or more than the one required by the deterministic Grover's algorithm (the one based on generalized Householder reflections). We use various semiempirical methods to show that the interval of number of iterations for which the algorithm keeps the probability of finding solution high depends on the register size and the oracle phases.*

## 1. Introduction

There are various quantum algorithms for searching for an element in an unordered database, examples include Grover's search [1], quantum random walk search [2] and fixed-point quantum search [3]. Each of the three algorithms mentioned is quadratically faster than the classical search algorithms and has applications for which it is more suitable than the other two. Grover's algorithm requires a smaller quantum register and fewer oracle calls than the other two algorithms. It can be constructed easily on various physical systems that can efficiently implement generalized Householder reflections such as ion traps [4] and systems based on photonic qubits [5]. The original Grover's algorithm is probabilistic, however there are various modifications that can be made to it to make it deterministic [6] [7] [8]. The one that uses generalized Householder reflections and phase matching [6] is suitable for construction in these quantum computer architectures. These properties make this algorithm very attractive for theoretical study and experimental implementations on various physical systems.

Grover's algorithm is also an important subroutine in various other subfields of quantum computing. The most notable algorithm that uses Grover's search as a subroutine is the quantum counting algorithm [9], used for finding the number of solutions of Grover's and the quantum random walk search algorithms. Other algorithms that uses Grover's search are: the quantum algorithm for the vertex coloring problem [10], algorithms for solving image pattern matching [11], collision detection algorithm [12] and solving various satisfiability problems [13]. Grover's algorithm also has applications in quantum machine learning – for support vector machines [14], speeding up neural network training [15] and quantum genetic algorithms [16]. Applications of Grover's search as a component in quantum cryptographic protocols include a quantum key distribution protocol [17], a quantum dialogue protocol [18] and two quantum secret sharing protocols [19] [20].

In statistical modeling, regression analysis is used to estimate the correlation between two or more variables. For a proper description of processes with various kinds of behavior, a variety of functions and parametric curves are used such as the Hill function [21] and superellipses [22].

The Hill function is a nonlinear logistic regression function [23] has a wide variety of applications in systems biology [24], pharmacology [25], mathematical and computational biology [26], where it is used to describe nonlinear processes. It's plot is similar to that of the sigmoid function, but its slope is controllable and it can be modified to fit Gaussian like functions. Modifications of this function were used to study various modifications of Grover's [27] and quantum random walk search [28] algorithms.

A superellipse (known also as a Lame curve) is a parametric curve, that is used for fitting functions in a variety of fields, including the characteristic equation of photovoltaics [29], face recognition in computer vision [30], network traffic classification in supervised machine learning [31], pattern generation and face generation in computer graphics [32] [30] and modeling various shapes in biology [33].

The current work is structured as follows: In Section 2, a short review of Grover's search algorithm is provided, as well as a description of some of its modifications. In Subsection 2.1. we give a brief description of a variant of Grover's algorithm that uses qudits. Subsection 2.2. shows the algorithm's geometric interpretation and its connection to the required number of iterations of the algorithm. Subsection 2.3. examines a modification of the algorithm that uses generalized Householder reflections and shows a way to use this modification to make the algorithm deterministic. Subsection 2.4. describes a modification of Grover's algorithm that uses multiphase oracle. Our new results are presented in Section 3. A mathematical description of our model for the general case of multiphase oracle, as well as analytical calculations for the Grover's operator, are provided in Subsection 3.1. After that, we focus on the simplest case when there are only two solutions marked with different phases by the oracle. In Subsection 3.2. and Subsection 3.3. we give the results of Monte Carlo simulations for the probability of finding a solution as a function of the two marking phases of the oracle. In particular, we focus on the oracle phase values for which a sufficiently high probability of finding a solution is reached after a number of iterations not greater than the optimal one for the standard Grover's algorithm plus one, and show that they can be fitted using a quarter of superellipse. In Subsections 3.3. and 3.5. we use semiempirical methods to find the superellipse parameters. One such method is reviewed in Subsection 3.4. In Subsection 3.6. we give an estimation of the phases that lead to maximal robustness against decreasing the probability after surpassing the number of iterations required by the standard Grover's algorithm as a function of the register size as well as an evaluation of this robustness. In Subsection 3.7. the cases for a larger number of phases are discussed. We make a brief summary of our results in the Conclusion (Section 4).

## 2. Grover's algorithm and some of its modifications

Grover's search algorithm is a probabilistic quantum algorithm for searching in an unordered database and returns one of the elements fulfilling the search conditions. These elements are also called solutions. Invented in 1996 [1], it is one of the oldest quantum algorithms and is quadratically faster than the best known classical algorithms. Together with Shor's factoring algorithm they show that quantum computers can efficiently solve problems that a classical computer would not be able to solve within a feasible amount of time and/or memory.

### 2.1. Grover's algorithm with qudits

Grover's algorithm uses a single quantum register whose size is equal to the number $N$ of the elements in the database. Let $M$ be the number of solutions, the final goal is to find one of those $M$ elements. Here we give a brief description of the algorithm's procedure:

In the beginning of the algorithm, the register is in the state $|0\rangle_N$. It is then put in an equal superposition $|\psi(0)\rangle_N$ of states. One way to do this is by using a quantum Fourier transformation matrix $F$.

$$|\psi(0)\rangle_N = F|0\rangle_N = \frac{1}{\sqrt{N}} \sum_{j=0}^{N-1} |j\rangle_N \qquad (1)$$

Then, the Grover iteration is applied to the register $t_{iter}$ times. The iteration consists of the following gates:

1) An oracle i. e. a reflection against the state that is an equal superposition of all non-solution states. It is based on a function that can recognize the solution states, and marks them by flipping their sign. Let us denote the solution states by $|m_i\rangle_N$ with $(j = 0,1, \dots, M-1)$. The oracle can be expressed as:

$$O_N(m_i) = I_N - 2|\beta\rangle_N\langle\beta|_N \qquad (2)$$

where $I_N$ is an identity matrix of dimension $N$ and $|\beta\rangle$ is the equal superposition of all solution states.

2) A reflection operator against the equal superposition of all states:

$$P_N = I_N - 2|\psi(0)\rangle_N\langle\psi(0)|_N \qquad (3)$$

Let us denote the whole Grover iteration by $G_N$, thus:

$$G_N = P_N O_N \qquad (4)$$

After the required number of iterations, a measurement is done on the register. The probability that the outcome of the measurement is a solution state is high and depends on both the register size and the number of solutions. The quantum circuit of Grover's algorithm is shown in $Fig.1$.

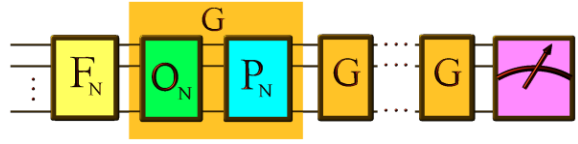

*Fig.1. Quantum circuit of Grover's search algorithm. Each of the quantum gates is shown in a different colour: the Fourier transform $F_N$ operator in yellow, the oracle $P_N$ in green, the reflection operator $P_N$ in teal and measurement in magenta. The Grover's iteration $G_N$ is shown with an orange rectangle around the gates $O_N$ and $P_N$.*

It is important for our goals to note that the probability of finding a solution is a sine like periodic function of the number of iterations. This means that after the required number of iterations is surpassed, the probability of finding a solution starts decreasing.

*Fig. 2* a) shows the probability of obtaining each solution state after the required number of iterations in the case of register size $N = 200$ and solution states: $|12\rangle_{200}$, $|101\rangle_{200}$ and $|152\rangle_{200}$. The probability of finding a solution after each iteration in this case is shown on *Fig. 2* b).

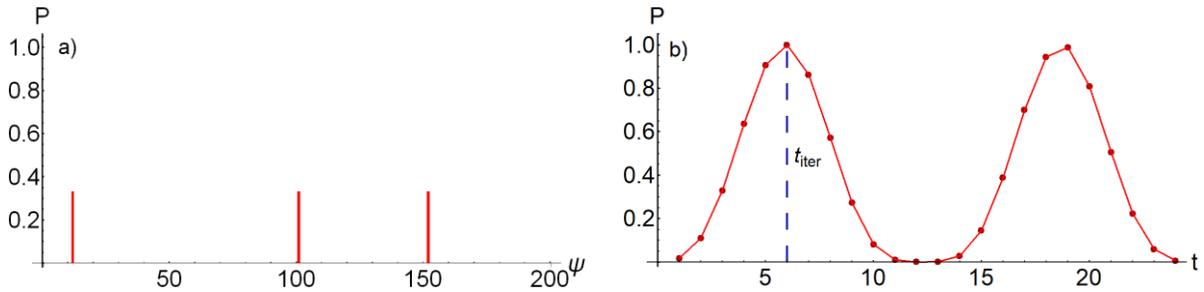

*Fig. 2. Probability of measuring each state in the end of the algorithm (plot a)) and the probability of obtaining a solution depending on the number of iterations (plot b)) for register size $N = 200$ and solution states $|12\rangle_{200}$, $|101\rangle_{200}$ and $|152\rangle_{200}$.*

## 2.2. Geometric interpretation and required number of iterations

Grover's iteration consists of two consecutive reflections, which result in a rotation in the space spanned between the equal superposition $|\beta\rangle_N$ of all solutions and the equal superposition $|\alpha\rangle_N$ of all elements that are not solutions. If $|m_j\rangle_N$ with $(j = 0, \ldots, M - 1)$ are the solution states and $|n_j\rangle_N$ with $(j = 0, \ldots, N - M - 1)$ are the non-solution states, then:

$$|\alpha\rangle_N = \frac{1}{\sqrt{N-M}} \sum_{i=0}^{N-M+1} |n_j\rangle_N \qquad (5)$$

$$|\beta\rangle_N = \frac{1}{\sqrt{M}} \sum_{j=0}^{M} |m_j\rangle_N \qquad (6)$$

Each iteration leads to rotation of the register state vector on an angle $\theta$. This angle depends on both the register size and the number of solutions:

$$\theta = 2 arcsin(\sqrt{M/N}) \qquad (7)$$

The initial state can be expressed as:

$$|\psi(0)\rangle_N = \cos\left(\frac{\theta}{2}\right)|\alpha\rangle_N + \sin\left(\frac{\theta}{2}\right)|\beta\rangle_N \qquad (8)$$

After the $t$-th iteration the register state becomes:

$$(G_N(M))^t|\psi(0)\rangle_N = \cos\left(\frac{2t+1}{2}\theta\right)|\alpha\rangle_N + \sin\left(\frac{2t+1}{2}\theta\right)|\beta\rangle_N \qquad (9)$$

For a maximal probability of finding a solution, the final state needs to be as close as possible to $|\beta\rangle_N$. This can be achieved if $\cos((2t_{iter}+1)\theta/2) \cong 0$. So, for the required number $t_{iter}$ of iterations we have:

$$\frac{\pi}{2} \cong \left(t_{iter} + \frac{1}{2}\right)\theta \qquad (10)$$

$t_{iter}$ must be an integer, so it is equal to the smallest integer that fulfills the equation (10).

$$t_{iter} = \left\lceil \frac{arccos[\sqrt{M/N}]}{2arcsin[\sqrt{M/N}]} \right\rceil \qquad (11)$$

where ⌈ ⌉ denotes rounding up the value in the brackets. When $M \ll N$ the following approximation can be applied:

$$t_{iter} \cong \left\lceil \frac{\pi}{4}\sqrt{N/M} \right\rceil \qquad (12)$$

For Grover's algorithm to be usable in database search, the number of solutions needs to be known. This number can be found by using quantum counting algorithm.

### 2.3. Grover's algorithm with generalized Householder reflections

For Grover's algorithm to be made deterministic, the rotation angle $\theta$ need to be reduced so that:

$$\frac{\pi}{2} = \left(t_{iter} + \frac{1}{2}\right)\theta \qquad (13)$$

One way to reduce the rotation angle is to construct the operators $O_N$ and $P_N$ using generalized Householder reflections.

$$O_N(\forall m_j, \varphi) = I_N - (1 - e^{i\varphi})|\beta\rangle_N\langle\beta|_N \qquad (14)$$

$$P_N(\omega) = I_N - (1 - e^{i\omega})|\psi(0)\rangle_N\langle\psi(0)|_N \qquad (15)$$

In this case, for the Grover iteration we have:

$$G_N(m_i, \varphi, \omega) = O_N(\forall m_j, \varphi) P_N(\omega) \tag{16}$$

The rotation angle $\theta$ depends on the phases of both operators. It can be shown [34] that for Grover's algorithm to be deterministic, the two phases need to be equal $\varphi = \omega = \varphi_{max}$. The Grover iteration achieves maximum rotation when $\omega = \pi$ and in this case it is reduced to a standard Householder reflection. The minimal rotation is achieved for $\omega = 0$, when it is reduced to an identity operator. The value of the optimal phase depends on the number of solutions and the register size. It can be evaluated by the formula:

$$\varphi_{max} = 2\arcsin\left(\sin\left(\frac{\pi}{6 + 4\left\lfloor\frac{1}{4}\left(-2 + \frac{\pi}{\arcsin(\sqrt{M/N})}\right)\right\rfloor}\right)\sqrt{N/M}\right) \tag{17}$$

where $\lfloor\ \rfloor$ denotes rounding down the value in the brackets.

**2.4. Grover's algorithm with multi-phased oracle for searching ranked targets**

Using a Grover's search with an oracle that marks each solutions with different phase was recently introduced in [35]. They use those phases in order to change the probability to find each solution at the end of the algorithm's implementation. Such modification can be used when finding each solution has different priority.

Let there is a quantum register of size $N$ with $M$ solution states $|m_j\rangle_N$ ($j = 0,1,\ldots,M-1$). On each $|m_j\rangle_N$, a phase $\varphi_j$ is applied, non-solution states $|n_h\rangle_N$ ($h = 0,1,\ldots,N-M-1$) remain unmodified by the oracle. Let the states of the register $|\psi(t)\rangle_N$ after the $t$-th iteration be:

$$|\psi(t)\rangle_N = \sum_{j=0}^{M-1} \beta_j(t)|m_j\rangle_N + \alpha(t) \sum_{h=0}^{N-M-1} |n_h\rangle_N \tag{18}$$

Then after the oracle $O_N(\forall m_j, \forall \varphi_j)$ is applied to it, the register state becomes:

$$O_N(\forall m_j, \forall \varphi_j)|\psi(t)\rangle_N = \sum_{j=0}^{M-1} \beta_j(t)e^{i\varphi_j}|m_j\rangle_N + \alpha(t) \sum_{h=0}^{N-M-1} |n_h\rangle_N \tag{19}$$

Note that all non-solution states evolve in the same way during the algorithm.

The oracle $O_N(\forall m_j, \forall \varphi_j)$ is equivalent to M consecutive oracles $O_N(m_j, \varphi_j)$, each of which marks the corresponding solution with its corresponding phase $\varphi_j$, thus:

$$O_N(\forall m_j, \forall \varphi_j) = \prod_{i=0}^{M-1} O_N(m_j, \varphi_j) \tag{20}$$

where:

$$O_N(m_j, \varphi_j) = I_N - (1 - e^{i\varphi_j})|m_j\rangle_N \langle m_j|_N \qquad (21)$$

As second reflection they used a standard householder reflection operator:

$$P_N = I_N - 2|\psi(0)\rangle_N \langle \psi(0)|_N \qquad (22)$$

The final state is obtained after applying the Grover's iteration $t_{iter}$ times:

$$|\psi(t)\rangle_N = (P_N O_N)^{t_{iter}} |\psi(0)\rangle_N \qquad (23)$$

### 3. Oracle applying different phases to each solution

The original Grover's algorithm has a periodic behavior as a Grover iteration can be seen as a rotation in the space spanned by two vectors, one of which is an equal superposition of all solutions and the other one is an equal superposition of all non-solutions. When the required number of iterations to obtain the solution is surpassed, the probability of obtaining it decreases as fast as the algorithm finds it. Here we present a modification of Grover's algorithm for the case of more than one solution that keeps the probability of finding a solution close to the maximal one for more iterations than the standard Grover's algorithm. For the sake of simplicity of presentation we will call this characteristic - robustness against more iterations.

#### 3.1. General case

Our quantum register size $N$, with $M$ solution states $|m_j\rangle_N$ ($j = 0,1, \dots, M-1$). The oracle applies on each $|m_j\rangle_N$, a phase $\varphi_j$ and non-solution states $|n_h\rangle_N$ ($h = 0,1, \dots, N-M-1$) remains the same. We use the same oracle as the one used in [35]:

$$O_N(\forall m_j, \forall \varphi_j) = \prod_{i=0}^{M-1} \left( I_N - (1 - e^{i\varphi_j})|m_j\rangle_N \langle m_j|_N \right) \qquad (24)$$

The Grover operator $G_N$ is constructed using $O_N(\forall m_j, \forall \varphi_j)$ and for $P_N$ is used generalized Householder reflection with phase $\omega$:

$$P_N(\omega) = I_N - (1 - e^{i\omega})|\psi_0\rangle_N \langle \psi_0|_N \qquad (25)$$

Let's define the state $|\psi(t)\rangle_N$ of the register after the $t$-th iteration as:

$$|\psi(t)\rangle_N = \sum_{j=0}^{M-1} \beta_j(t)|m_j\rangle_N + \alpha(t) \sum_{h=0}^{N-M-1} |n_h\rangle_N \qquad (26)$$

At each iteration of the algorithm, $G_N$ is applied on the state of the register:

$$|\psi(t+1)\rangle_N = P_N O_N |\psi(t)\rangle_N = G_N |\psi(t)\rangle_N \qquad (27)$$

It can be shown that between two consecutive iterations, the following recursive relation exists:

$$\alpha(t+1) = \alpha(t)\frac{M + e^{i\omega}(N-M)}{N} - \frac{1}{N}(1 - e^{i\omega})\sum_{j=0}^{M-1} e^{i\varphi_j}\beta_j(t) \quad (28)$$

$$\beta_h(t+1) = \beta_h(t)e^{i\varphi_h} - \alpha(t)(1-e^{i\omega})\frac{N-M}{N} - \frac{1}{N}(1-e^{i\omega})\sum_{j=0}^{M-1} e^{i\varphi_j}\beta_j(t)$$

Since, as mentioned above, non-solution states evolve the same way during the proceeding of the algorithm it is more convenient to express the register state in the basis $\{|A\rangle_N, |m_1\rangle_N, \ldots, |m_M\rangle_N\}$, where $|A\rangle_N$ is an equal superposition of all non-solutions:

$$|A\rangle_N = \frac{1}{\sqrt{N-M}}\sum_{j=0}^{N-M-1} |n_j\rangle_N \quad (29)$$

We also have the following normalization condition for the register state vector:

$$(N-M)|\alpha(t)|^2 + \sum_{j=0}^{M-1}|\beta_j(t)|^2 = 1 \quad (30)$$

Thus, the Grover iteration can be expressed in the basis $\{|A\rangle_N, |m_1\rangle_N, \ldots, |m_M\rangle_N\}$ as:

$$G_N = \begin{pmatrix} \frac{M+e^{i\omega}(N-M)}{N} & -\frac{\sqrt{N-M}}{N}(1-e^{i\omega})e^{i\varphi_0} & \cdots & -\frac{\sqrt{N-M}}{N}(1-e^{i\omega})e^{i\varphi_{M-1}} \\ -\frac{\sqrt{N-M}}{N}(1-e^{i\omega}) & e^{i\varphi_0} - \frac{1}{N}(1-e^{i\omega})e^{i\varphi_0} & \cdots & -\frac{1}{N}(1-e^{i\omega})e^{i\varphi_{M-1}} \\ \vdots & \vdots & \ddots & \vdots \\ -\frac{\sqrt{N-M}}{N}(1-e^{i\omega}) & -\frac{1}{N}(1-e^{i\omega})e^{i\varphi_0} & \cdots & e^{i\varphi_{M-1}} - \frac{1}{N}(1-e^{i\omega})e^{i\varphi_{M-1}} \end{pmatrix} \quad (31)$$

The evolution of the system strongly depends on the register size, the number of solutions and the corresponding phases $\omega, \varphi_0, \ldots, \varphi_{M-1}$. The evolution operator can be decomposed into Givens rotations. In the general case the composition is very complex, however when $\omega = \pi$, the result simplifies significantly. The decomposition consists of standard rotations between $|A\rangle_N$ and $|m_i\rangle_N$ and complex rotation. This leads to a more complex corelation between the probability of finding a solution on the phases.

An example for the probability $P$ of obtaining a solution after each iteration $t$ of the algorithm in the case of two solutions and a register size of $N = 200$ is shown on Fig. 3. with a solid red line. Each plot corresponds to various values of the phases $\varphi_0$ and $\varphi_1$ which are also shown on that plot. The probabilities of finding the solutions marked with those phases are shown with a dashed green and dashed teal line respectively. The vertical dashed blue line corresponds to the required number of iterations of the deterministic Grover's algorithm with two equal and optimal phases.

The dependence of the probability obtaining a solution on the number of current iterations is nonlinear and strongly depends on the values of the phases. For this reason, in the current paper we will study the robustness against a large number of iterations numerically. We will only focus on the probability of finding a solution in the simplest case when there are only two solutions, each marked with a different phase.

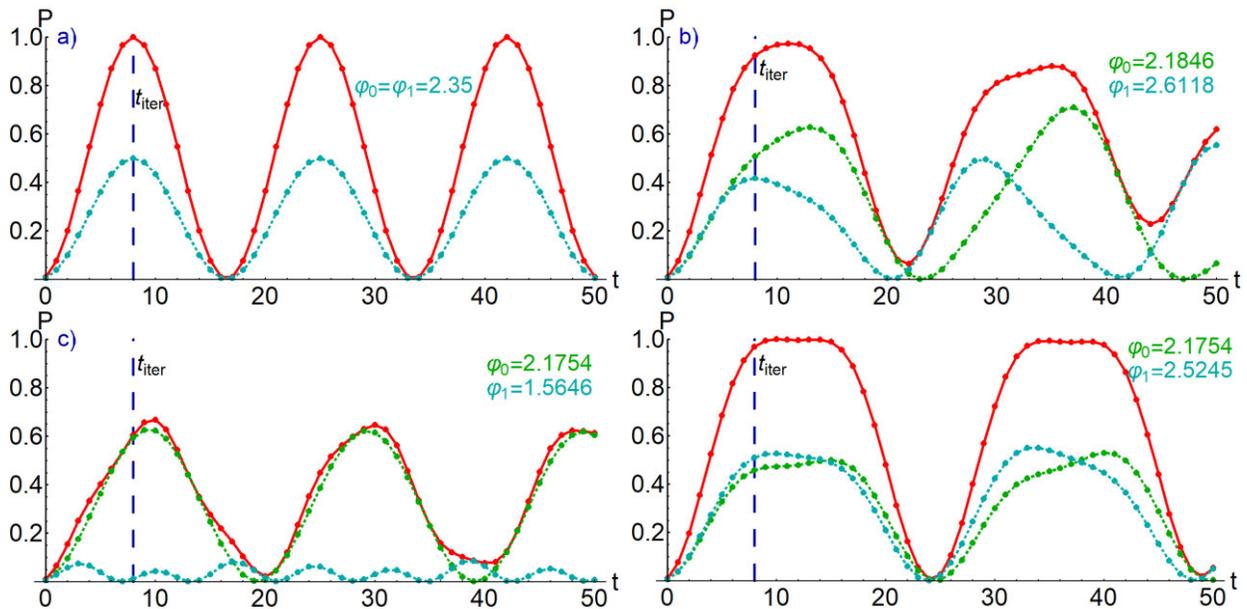

Fig. 3. Probability of finding a solution of Grover's algorithm with an oracle applying two different phases as a function of the number of iterations, for register size $N = 200$ and a phase of the reflection operator $P_{200}$ equal to the optimal oracle phase $\varphi_{max} \cong 2.34997$ in the deterministic Grover's algorithm. Each plot corresponds to different value of the oracle phases $\varphi_0$ and $\varphi_1$. The dark green and the teal dashed line correspond to the probability of finding the solution marked by the oracles with phases $\varphi_0$ and $\varphi_1$ respectively. The solid red line shows the probability of obtaining any solution and the dashed blue line shows the required number of iterations for Grover's algorithm with two solutions and equal phases.

### 3.2. Monte Carlo simulations

For a combination of phase values $\varphi_0$ and $\varphi_1$ to give a useful Grover's algorithm, the following requirements must be fulfilled:

1) The maximal probability of finding a solution must not be lower than in the original Grover's algorithm;
2) It must need the same number of iterations to obtain a solution as the original Grover's algorithm with two marked states

We did Monte Carlo simulations of the algorithm, during which the two phases $\varphi_0$ and $\varphi_1$ are chosen at random. The reflection $P_N$ is constructed using the optimal phase. At each simulation the following data is saved. The values of the phases, $\varphi_0$ and $\varphi_1$, the maximal probability $P_{max}(\varphi_0, \varphi_1)$ of finding a solution and the number $t_{iter}(\varphi_0, \varphi_1)$ of iterations required to achieve this probability.

Fig. 4 shows the results from the Monte Carlo simulations for register sizes N=200 (plot a)) and N=325 (plot b)). On the plots spanned between the $\varphi_0$ and $\varphi_1$ axes all points $(\varphi_0, \varphi_1)$ that satisfy the constraints: $P_{max}(\varphi_0, \varphi_1) > 0.92$ and $t_{iter}(\varphi_0, \varphi_1) < t_{iter}(\varphi_{max}, \varphi_{max}) + 2$ are shown with red dots. The plot of $\varphi_0 = \varphi_1$ is shown with a dashed green line.

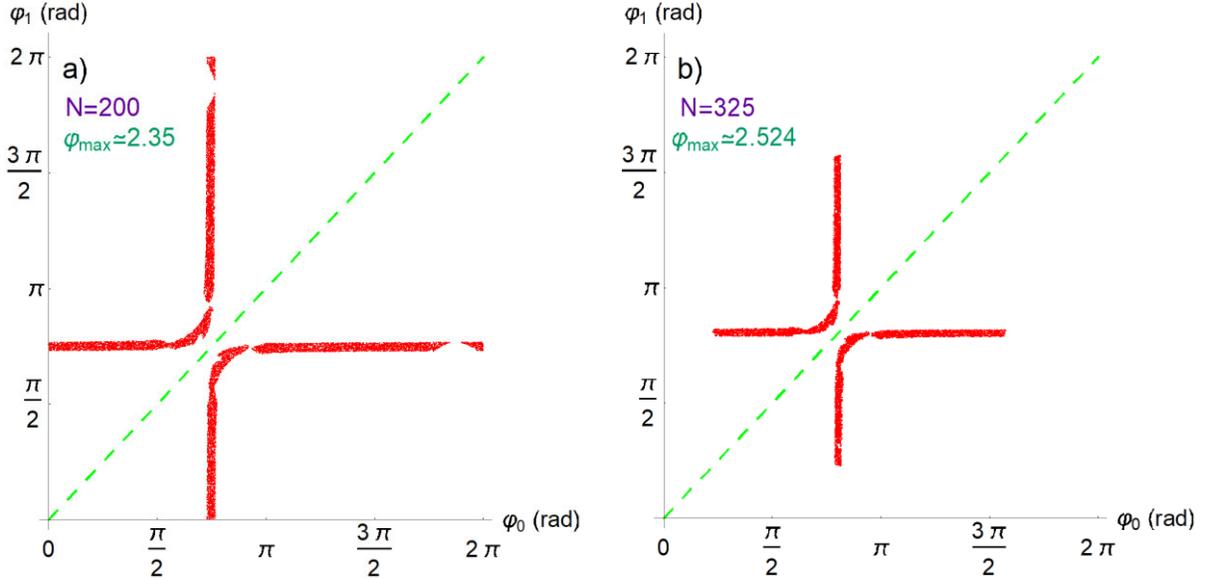

Fig. 4. Plots of the values of the oracle phases $\varphi_0$ and $\varphi_1$ from the Monte Carlo simulations of Grover's algorithm for which the modified Grover's algorithm with a multiphase oracle reaches a probability of finding a solution equal to or more than 0.92, for no more than one iteration more than the optimal number of iterations in the standard Grover's algorithm. These values are with red dots. The plot of $\varphi_0 = \varphi_1$ is shown with a green dashed line shows the line. Plots a) and b) show the results for register size $N = 200$ and $N = 325$ respectively.

From the simulations it can be seen that the points $(\varphi_0, \varphi_1)$ fulfilling the requirements form two stripes that get thinner with the increase in the register size. So a small deviation in the $\varphi_0$ and $\varphi_1$ values can lead to going out of the stripe. This shows that there is a limit of the applicability of our model.

It is important to note that the overall probability of finding a solution is invariant against exchanging the values of $\varphi_0$ and $\varphi_1$. This leads to the symmetry of the plots against the axis $\varphi_0 = \varphi_1$ so it will be enough to study only one of the two stripes. We chose to study the stripe for which $\varphi_0 > \varphi_1$. This stripe can be approximated with a parametric curve – the top left quadrant of a superellipse translated along the $\varphi_0$-axis.

### 3.3. Superellipse fit

A superellipse has the following parametric equation:

$$\begin{cases} \varphi_0 = s_{\varphi_0} + a_{\varphi_0} \cdot |\cos(z)|^{2/p_\varphi} \\ \varphi_1 = a_{\varphi_1} \cdot |\sin(z)|^{2/p_\varphi} \end{cases} \quad z \in [0, 2\pi] \quad (32)$$

In order to find the values of $\varphi_0$ and $\varphi_1$ that lead to highest robustness against a larger number of iterations, we need to:

1) Find all parameters of the fitting curve and how they change with the increase of the register size
2) Find the part of the parametric curve that gives the highest robustness

The point where the curve crosses the $\varphi_0$ axis ($\varphi_0 = \varphi_{max}, \varphi_1 = 0$) corresponds to Grover's algorithm with only one solution. We also want to obtain the top left quadrant of the superellipse. These two restrictions are fulfilled when $s_{\varphi_0} = 2\pi$ and $a_{\varphi_0} = \varphi_{max} - 2\pi$. Grover's algorithm with only one solution can also be obtained using $\varphi_0 = 2\pi$, and $\varphi_1 = \varphi_{max}$, so we find $a_{\varphi_1} = \varphi_{max}$.

The parameter $p_\varphi$ need to be found numerically. This parameter depends on the register size $N$, so we must find a semiempirical equation that can be used for a wide range of register sizes. It is important to note that a relatively small share of the points in the Monte Carlo simulation fulfill the requirements 1) and 2) in Subsection 3.2., so there is significant error in the fitting of the parameter $p_\varphi$. On the other hand, the length of the superellipse's arc changes relatively slowly with the change of $p_\varphi$. For those reasons, we will use a value rounded to the first digit after the decimal point for all values for $p_\varphi$.

Numerical simulations were done for various register sizes. Two examples are shown on *Fig. 5*. The red dots represent the results of the Monte Carlo simulations when $\varphi_0 > \varphi_1$ and the blue line is the fitting curve obtained by using Eq. (32). The plots on the left and the right correspond to register sizes $N = 200$ and $N = 325$ respectively. The corresponding value of $p_\varphi$ used is also shown on each plot.

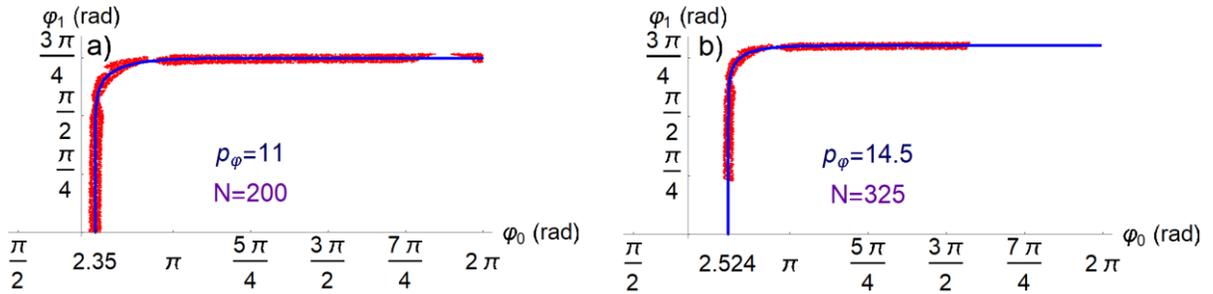

*Fig. 5. Comparison between the points that fulfill the conditions presented in this section (red dots) and the fitting superellipse curve (solid blue line) for register sizes $N = 200$ (plot a)) and $N = 325$ (plot b)).*

Our results show that a superellipse with these parameters fits the red points well. The disadvantage of this method is that for computational reasons we were unable to do these simulations for large register sizes. To make a generalization we will fit the results for $p_\varphi$ and use the resulting equation to make an extrapolation for larger values of $p_\varphi$.

The plot on *Fig. 6.* shows (with blue dots). the values of the fitting parameter $p_\varphi$ for various values of the register size $N$ The solid green line on the picture shows the fitting curve for these points.

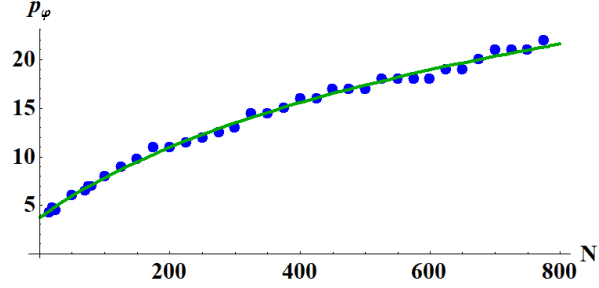

*Fig. 6. Plot showing the value of the parameter $p_\varphi$ from Eq. (32) for various register sizes (blue dots) and the fitting curve for it (green line).*

The function used to fit the blue dots on *Fig. 6.* is:

$$p_\varphi(N) = -65.7376 + 12.5476 * \text{Log}_e[N + 252.0719] \tag{33}$$

In the next subsection we define a semiempirical function that will be used to obtain a quantitative description of the robustness and to compare the robustness against more iterations of the Grover's algorithm for various register sizes and values of the phases $\varphi_0$ and $\varphi_1$.

### 3.4. Asymmetric modified Hill function – definition and example

In order to compare numerically, we introduce the function that we will use in order to compare the width of the curves – the asymmetric modified Hill functions:

$$W(t, b, k_L, n_L, k_R, n_R, w) = \frac{b}{2}\left(\frac{([\![t-w]\!]_\pm + 1)k_R^{n_R}}{|t-w|^{n_R} + k_R^{n_R}} - \frac{([\![t-w]\!]_\pm - 1)k_L^{n_L}}{|t-w|^{n_L} + k_L^{n_L}}\right) \tag{34}$$

where $[\![t-w]\!]_\pm$ means the sign of $t-w$.

This function is an asymmetric analogue of the modified Hill function described in [28]. The parameters $n_R$ and $n_L$ correspond to the slope of the curve on the right and the left respectively, $k = k_L + k_R$ correspond to the width of the plateau, $b$ gives the maximal height, $w$ is the point where the two branches of the function meet.

*Fig. 7.* shows plots of the function $W(t, b, k_L, n_L, k_R, n_R, w)$ for various parameter values. On each plot $W(t, 3, 1.7, 5, 1.1, 3, \pi)$ is shown with dotted teal line. The solid red and dashed green lines show the curves $W$ for which all parameters except one have the same

values as for the teal line. On plot a) the solid red and the dashed green line correspond to $b = 2$ and $b = 2.5$ respectively. On plot b) they correspond to $k_L = 1$ and $k_L = 1.3$ respectively, on plot c) to $n_L = 1$ and $n_L = 1.3$ respectively, on plot d) to $k_R = 1.4$ and $k_R = 2$ respectively on plot e) to $n_R = 5$ and $n_R = 33$ respectively and on f) to $w = \pi/4$ and $w = \pi/2$.

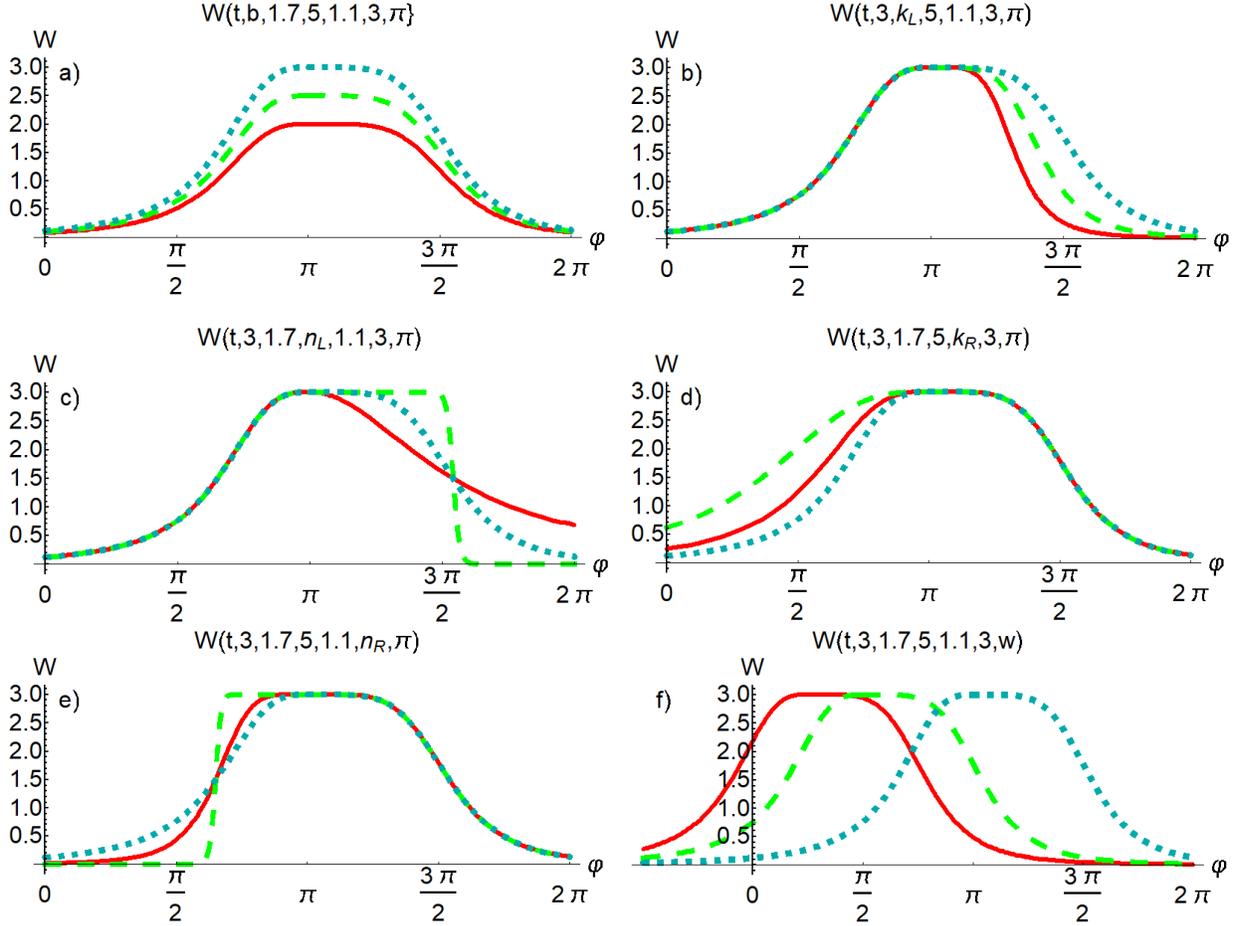

Fig. 7. Plots of $W(t, b, k_L, n_L, k_R, n_R, w)$ as defined by Eq. (34) as a function of $t$ where $b, k_L, n_L, k_R, n_R$ and $w$ are parameters. Each plot features three plots with different colors and dashing corresponding to a different value of one of the parameters and fixed value of the other parameters (those values are shown in the corresponding plot's captions. On plot a) the solid red, dashed green and teal dotted lines correspond to b=2; b=2.5 and b=3 respectively. On plot b) they correspond to $k_L = 1$, $k_L = 1.3$ and $k_L = 1.7$ respectively; on plot c) to $n_L = 2$, $n_L = 45$ and $n_L = 5$ respectively; on plot d) to $k_R = 1.4$, $k_R = 2$ and $k_R = 1.1$ respectively; on plot e) $n_R = 5$, $n_R = 33$ and $n_R = 3$ respectively; and on f) $w = \pi/4$, $w = \pi/2$ and $w = \pi$ respectively.

An estimation of how good each particular fit is, can be done by evaluating the standard deviations:

$$\sigma = \sqrt{\sum_{j=1}^{\varrho} \frac{\left(W(t, b, k_L, n_L, k_R, n_R, w) - P(t, N, \varphi_0, \varphi_1)\right)^2}{\varrho - q}} \tag{35}$$

Here $P_j(t, N, \varphi_0, \varphi_1)$ is the probability of obtaining a solution at the $t$-th iteration for a register size of $N$ and oracle phases $\varphi_0$ and $\varphi_1$. $W(t, b, k_L, n_L, k_R, n_R, w)$ is the fit with the asymmetric modified Hill function and its parameters, $\varrho$ is the number of points used for the fitting and $q$ is the number of fitting parameters (in the case of $W$, $q = 6$).

### 3.5. Semiempirical method to study the maximal robustness

Asymmetric modified Hill functions can be used to fit the probability $P(t, \varphi_0(z), \varphi_1(z))$ of obtaining a solution, for various values of the phases $\varphi_0(z)$ and $\varphi_1(z)$. Such a quantitative description gives us the possibility to compare the curves for $P(t)$ for various values of $z$.

On *Fig. 8* shows two such fits, for register size $N = 200$ and $M = 2$ solutions. For both of them Eq. (32) is used to obtain both phases ($\varphi_0(z)$ and $\varphi_1(z)$) for the oracle. The plots on the left and the right corresponds to $z = 2.651$ and $z = 4.172$ respectively. The vertical blue dashed line shows the oracle phase used in the deterministic Grover's algorithm.

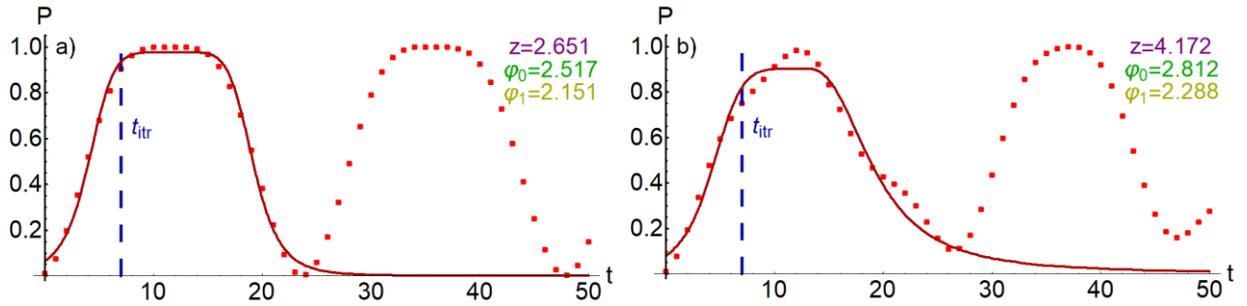

*Fig. 8. Plot of the probability of finding a solution (red dots) as a function of the number of iterations and the fitting modified Hill function curve for them (dark red line) for register size $N = 200$ and $M = 2$ solutions. Oracle phases $\varphi_0$ and $\varphi_1$ are evaluated using Eq. (32) for $z= 2.651$ (plot a)) and $z= 4.172$ (plot b)).*

From Fig (34) it can be seen that, the points can be fitted well with a asymmetric modified Hill function curve when they form a plateau were the probability of obtaining a solution is close to the maximal one. In the case where there is no such plateau, the value of the parameter $b$ becomes significantly lower. This is because the fitting averages the height of the curve that is approximated by the plateau. We denote the value of $b$ for particular value of $z$ and $N$ by $b(N, z)$.

Both $k_L$ and $k_R$ contribute to the width of the plateau of the fitting function. However, when $n_L$ is an even number, the result of the fitting does not depend on the sign of $k_L$. Similar is the case for $n_R$ and $k_R$. Thus, one of the parameters $k_L$ or $k_R$ or both of them can have a negative value. For this reason, for quantitative description of the robustness against more iteration we use the value of $k = |k_L| + |k_R|$. We denote the value of $k$ for particular values of $z$ and $N$ by $k(N, z)$.

Pictures on *Fig. 9* show the fitting for parameters $b(N, z)$ (on plot a)) and $k(N, z)$ (on plot b)) of the fitting modified Hill function for $P(N, t, \varphi_0(z), \varphi_1(z))$, for register size of $N = 200$ and $z \in [0, 2\pi]$.

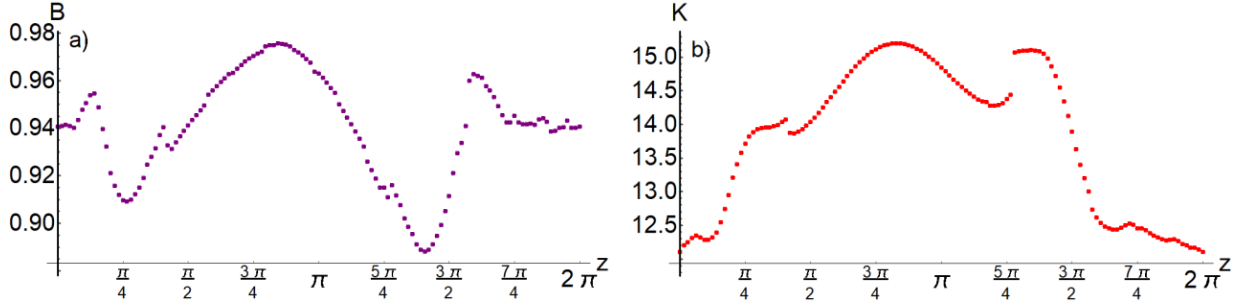

*Fig. 9. Plots of the values of the parameters b (plot a)) and k (plot b)) of the fitting modified Hill function for $P(N, t, \varphi_0(z), \varphi_1(z))$ as functions of the parameter z of the fitting superellipse function for $\varphi_0(z)$ and $\varphi_1(z)$ for register size of 200.*

It can be seen that the value of $z$ that gives the highest robustness against more iterations cannot be obtained using either of the parameters $b(N, z)$ an $k(N, z)$ alone. For this reason, we introduce the following quantity:

$$\Omega(N, z) = \frac{b(N, z) . k(N, z)}{b_{max}(N) . k_{max}(N)} \tag{36}$$

where $b_{max}(N)$ and $k_{max}(N)$ are the maximal values of $b$ and $k$, for the same values of $N$ and all possible $z \in [0, 2\pi]$.

The result shows that this quantity always gives good results for the value of $z$ for which the algorithm is the most robust against a larger number of iterations. *Fig. 10* show examples for $\Omega(N, z)$ for $N = 200$ (plot a)) and $N = 325$ (plot b)). Plots c) and d) show the results of the numerical simulations of the Grover's algorithm for the corresponding optimal values of $z$ obtained for $N = 200$ and $N = 325$ respectively.

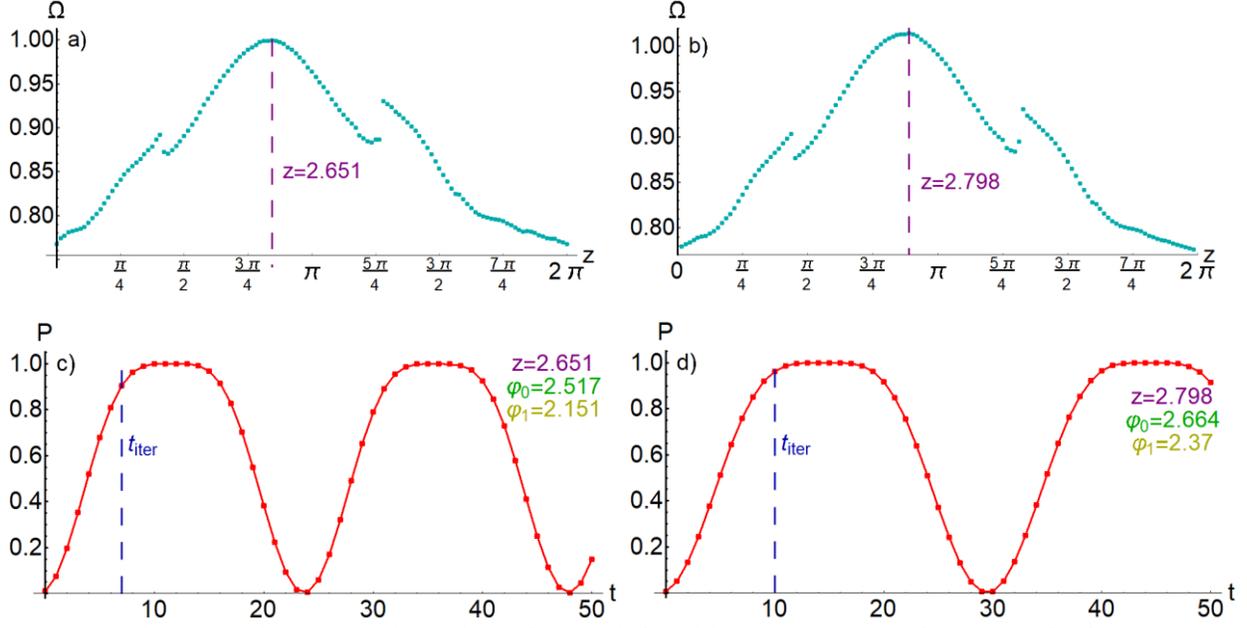

*Fig. 10. Plot a) and b) show the value of Ω as defined by Eq. (36) as a function of z for register sizes N=200 (plot a)) and N = 325 (plot b)). The dashed purple line shows the value of z that gives the highest value of Ω. Plot c) shows the result of the simulation of the modified Grover's algorithm with oracle phase values $\varphi_0$ and $\varphi_1$ calculated using Eq. (32) for register size N = 200 and the optimal value of z from plot a). Plot d) shows the results of the simulation of the algorithm for N = 325 and the optimal value of z from plot b).*

Let us denote by $z_{max}(N)$ the value of $z$ for a given register size $N$ that gives optimal robustness:

$$z_{max}(N) = z(N) \Leftrightarrow \Omega(N,z) = \Omega_{max}(N) \qquad (37)$$

where $\Omega_{max}(N)$ is the maximal value of $\Omega(N,z)$ for a fixed $N$.

This quantity is crucial for finding the optimal phases of $\varphi_0(z), \varphi_1(z)$ for a register of arbitrary size. The results of the numerical simulations for $z_{max}(N)$ for various register sizes from 20 to 775 is shown on *Fig. 11* with a teal solid line. On the same plot, the phase $\varphi_{max}(N)$ of the deterministic Grover's algorithm is shown with a dashed orange line for comparison.

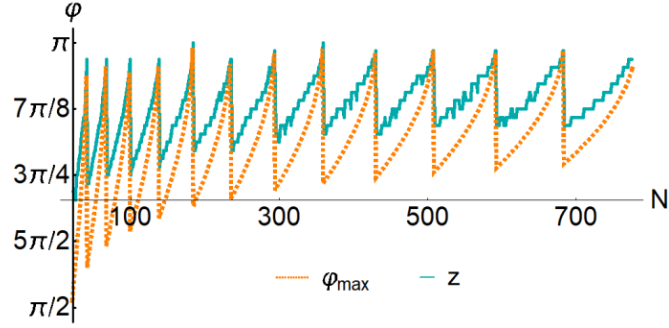

Fig. 11. Comparison between the value of the fitting parameter $z_{max}$ from Eq. (37) and the phase of the deterministic Grover's algorithm $\varphi_{max}$. The orange dashed line and the solid teal line shows the plots of $\varphi_{max}$ and $z_{max}$ respectively as functions of the register size.

It can be seen that $\varphi_{max}(N)$ and $z_{max}(N)$ have the same periodicity. At the values of $N$ just before the required number of iterations increases by one, the value of $z_{max}$ is slightly higher than the value of $\varphi_{max}$. However, with the increase of the number of iterations the difference between $z_{max}$ and $\varphi_{max}$ at those points decreases, so for larger number of iterations we can assume that at those points they are equal. Their value reach local peaks just before the point where the required number of iterations increases, at this points these values drop. Both the periodicity and the information about $\varphi_{max}$ can be used to find the values of $z_{max}$ for larger registers numerically.

One method that can be used to shrink the interval of possible values of $z_{max}$ is the following:

First, we find the values $V$ for which $\varphi_{max}(V)$ reaches local peaks, i. e. those satisfying the equation:

$$\left\lceil \frac{arccos[\sqrt{M/(V+1)}]}{2arcsin[\sqrt{M/(V+1)}]} \right\rceil - \left\lceil \frac{arccos[\sqrt{M/V}]}{2arcsin[\sqrt{M/V}]} \right\rceil = 1 \qquad (38)$$

Let us denote with $V_+$ the largest $V < N$ that satisfies equation (38) and with $V_-$, the smallest $V > N$ satisfying this equation. Then we define:

$$\Phi_+(N) = \varphi_{max}(N = V_+) \qquad (39)$$

$$\Phi_-(N) = \varphi_{max}(N = V_- + 1) \qquad (40)$$

So, for $z_{max}(N)$ we have:

$$z_{max} \epsilon [\Phi_-(N), \Phi_+(N)] \qquad (41)$$

In the interval between $(V_- + 1)$ and $V_+$ both the values of $z_{max}$ and $\varphi_{max}$ tend to increases.

## 3.6. Evaluation of the optimal phases and robustness against surpassing the required number of iterations

The oracle phases that give optimal robustness against a larger number of iterations are $\varphi_{0,\max}(N) = \varphi_{0,\max}(z_{\max}(N), N)$ and $\varphi_{1,\max}(N) = \varphi_{1,\max}(z_{\max}(N), N)$ so approximations for these phases are:

$$\begin{vmatrix} \varphi_{0,\max}(N) \approx 2\pi + (\varphi_{\max}(N) - 2\pi).|cos(z)|^{2/p_\varphi(N)} \\ \varphi_{1,\max}(N) \approx \varphi_{\max}(N).|sin(z)|^{2/p_\varphi(N)} \end{vmatrix} \quad z \in [\Phi_-(N), \Phi_+(N)] \quad (42)$$

where $p_\varphi(N)$ is the superellipse parameter defined in Subsection 3.3.

*Fig. 12* shows the results for $\varphi_{0,\max}(N)$ (solid dark red line) and $\varphi_{1,\max}(N)$ (dash dotted golden line) together with the optimal phase $\varphi_{\max}(N)$ of the deterministic Grover's algorithm (dotted orange line). It can be seen that the values of $\varphi_{0,\max}(N)$ and $\varphi_{1,\max}(N)$ are close to $\varphi_{\max}(N)$ for each and have the same periodicity as it.

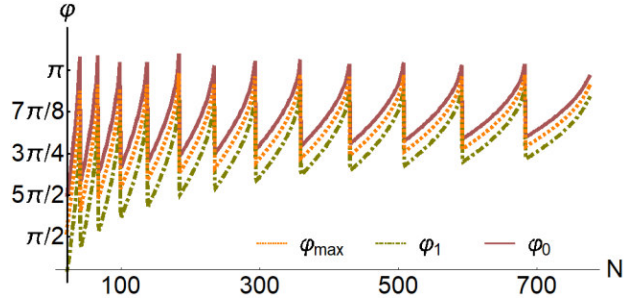

*Fig. 12. Optimal values $\varphi_{0,max}(N)$ (solid dark red line) and $\varphi_{1,max}(N)$ (dash-dotted golden line) of the oracle phases giving optimal robustness against a larger number of iterations, and the optimal oracle phase $\varphi_{max}(N)$ for the deterministic Grover's algorithm (dotted orange line)*

The value of $K_{\max}$ corresponds to the width of the plateau for the values of phases that give an optimal robustness. Based on the results, we can conclude that the robustness increases linearly with the increase of the natural logarithm of the register size. An approximate formula for $K_{\max}(N)$ is:

$$K_{\max}(N) \approx -82.658 + 15.978.|\ln[250.867 + N]| \quad (43)$$

*Fig. 13* shows the value of $K_{\max}$ (solid red line) as a function of the register size $N$, as well as a fitting curve (dotted purple line) obtained by using Eq. (43).

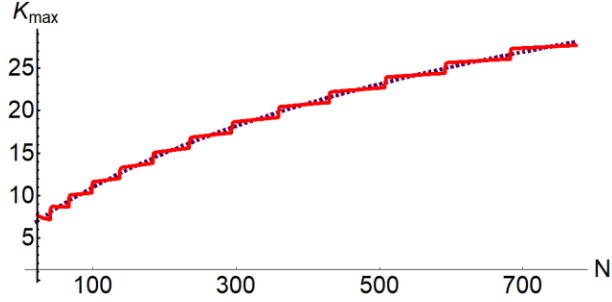

*Fig. 13. Values of the fitting parameter $K_{\max}(N)$ (solid red line) and an approximating curve (dotted purple line) obtained using Eq. (43).*

### 3.7. Discussion for the cases of more than two oracle phases

*Fig. 14* shows as red dots some results of our Monte Carlo simulations for the probability of finding a solution as a function of the number of iterations for a register size of 200 in the cases of three (picture a)) or four (picture b)) oracle phases. The values of the phases used are indicated on each plot.

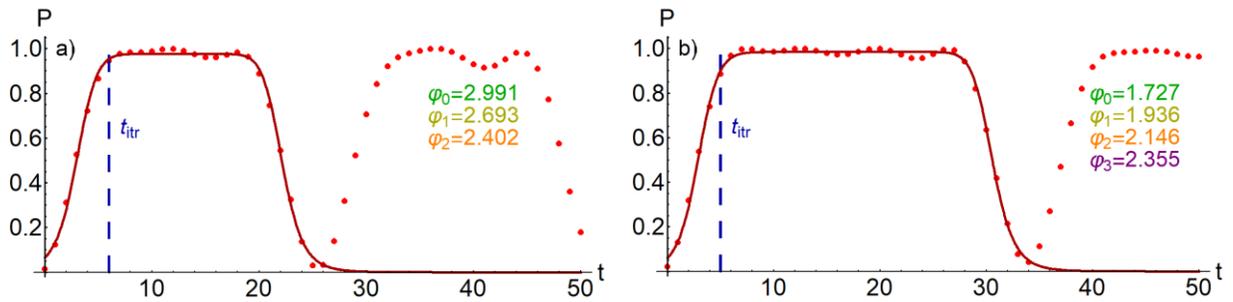

*Fig. 14. Probability of finding a solution (red dots) as a function of the number of iterations and the fitting asymmetrical modified Hill function curve for it (solid dark red line) for register size 200 in the case of three (plot a)) and four (plot b)) oracle phases. The phases are indicated on each plot.*

Our numerical simulations suggest that using higher number of oracle phases can lead to even higher robustness against a larger number of iterations. However, fitting the points that fulfil the conditions defined in Section 3.2. is much more complicated. Such fitting requires a parametric hypersurface of dimension equal to the number of oracle phases.

### 4. Conclusion

In this work we study a modification of the Grover's algorithm, with an oracle that marks each solution with different phase. We begin with defining the operators, and use their analytical form to obtain the general case for the Grover iteration and some useful recursive relations. In the general case the probability of finding a solution after each iteration depends on the register size, the number of solutions and the phases that the oracle applies to the solutions. We show examples that this modification can maintain a high probability of finding a solution even after the optimal number of iterations is surpassed, however, for this to be achieved, appropriate oracle phases need to be chosen. After that, we use numerical and semi analytical methods to study the case of only two solutions. Monte Carlo simulation are done to obtain the probability of finding a solution as a function of the oracle phases for various register sizes. We excluded all values of the phases for which the maximal probability of obtaining a solution is not high enough or which need more than one additional iteration to achieve this maximal probability and find out that the remaining points form stripes can be approximated with a quarter of superellipse. We introduce asymmetrical modified Hill functions in order to use them for fitting the probability of finding a solution as a function of the number of iterations. Using numerical simulations we observe how each fitting parameter depends on the register size. For the case of two solutions we obtain two formulas that can be used for extrapolations for our model. The first one is a semiempirical equation for the oracle phases that make the algorithm more robust against exceeding the number of iterations. The second one is a formula that evaluates how this robustness increases with the increasing of the register size. Finally we give some examples for such robustness for a larger number of phases.

## Acknowledgments

The work on this paper was supported by the Bulgarian National Science Fund under Grant KP-06-N58/5 / 19.11.2021.

## Acknowledgments

[1]     L. K. Grover, "A fast quantum mechanical algorithm for database search," in *Proceedings of the twenty-eighth annual ACM symposium on Theory of computing - STOC '96*, Philadelphia, Pennsylvania, United States: ACM Press, 1996, pp. 212–219. doi: 10.1145/237814.237866.

[2]     N. Shenvi, J. Kempe, and K. B. Whaley, "Quantum random-walk search algorithm," *Phys. Rev. A*, vol. 67, no. 5, p. 052307, May 2003, doi: 10.1103/PhysRevA.67.052307.

[3]     T. J. Yoder, G. H. Low, and I. L. Chuang, "Fixed-Point Quantum Search with an Optimal Number of Queries," *Phys. Rev. Lett.*, vol. 113, no. 21, p. 210501, Nov. 2014, doi: 10.1103/PhysRevLett.113.210501.

[4]     S. S. Ivanov, H. S. Tonchev, and N. V. Vitanov, "Time-efficient implementation of quantum search with qudits," *Phys. Rev. A*, vol. 85, no. 6, p. 062321, Jun. 2012, doi: 10.1103/PhysRevA.85.062321.

[5]     E. S. Kyoseva, D. G. Angelakis, and L. C. Kwek, "A single-interaction step implementation of a quantum search in coupled micro-cavities," *EPL*, vol. 89, no. 2, p. 20005, Feb. 2010, doi: 10.1209/0295-5075/89/20005.

[6]     G. L. Long, "Grover algorithm with zero theoretical failure rate," *Phys. Rev. A*, vol. 64, no. 2, p. 022307, Jul. 2001, doi: 10.1103/PhysRevA.64.022307.

[7]     G. Brassard, P. Hoyer, M. Mosca, and A. Tapp, "Quantum Amplitude Amplification and Estimation," vol. 305, 2002, pp. 53–74. doi: 10.1090/conm/305/05215.

[8]     P. Høyer, "Arbitrary phases in quantum amplitude amplification," *Phys. Rev. A*, vol. 62, no. 5, p. 052304, Oct. 2000, doi: 10.1103/PhysRevA.62.052304.

[9]     M. A. Nielsen and I. L. Chuang, *Quantum computation and quantum information*. Cambridge: Cambridge Univ. Press, 2007.

[10]    S. B. Mandal, A. Chakrabarti, and S. Sur-Kolay, "Synthesis of Ternary Grover's Algorithm," in *2014 IEEE 44th International Symposium on Multiple-Valued Logic*, May 2014, pp. 184–189. doi: 10.1109/ISMVL.2014.40.

[11]    H. Tezuka, K. Nakaji, T. Satoh, and N. Yamamoto, "Grover search revisited: Application to image pattern matching," *Phys. Rev. A*, vol. 105, no. 3, p. 032440, Mar. 2022, doi: 10.1103/PhysRevA.105.032440.

[12]    Y. Shi, "Quantum lower bounds for the collision and the element distinctness problems," in *The 43rd Annual IEEE Symposium on Foundations of Computer Science, 2002. Proceedings.*, Nov. 2002, pp. 513–519. doi: 10.1109/SFCS.2002.1181975.

[13]    R. Wang, "Comparing Grover's Quantum Search Algorithm with Classical Algorithm on Solving Satisfiability Problem," in *2021 IEEE Integrated STEM Education Conference (ISEC)*, Mar. 2021, pp. 204–204. doi: 10.1109/ISEC52395.2021.9764017.

[14]    D. Anguita, S. Ridella, F. Rivieccio, and R. Zunino, "Quantum optimization for training support vector machines," *Neural Networks*, vol. 16, no. 5, pp. 763–770, Jun. 2003, doi: 10.1016/S0893-6080(03)00087-X.

[15]    A. Haverly, S. Rahimi, and M. A. Novotny, "Grover's Algorithm for Neural Networks," in *2025 International Conference on Quantum Communications, Networking, and Computing (QCNC)*, Mar. 2025, pp. 667–671. doi: 10.1109/QCNC64685.2025.00111.

[16]    M. Udrescu, L. Prodan, and M. Vlăduţiu, "Implementing quantum genetic algorithms: a solution based on Grover's algorithm," in *Proceedings of the 3rd conference on Computing frontiers*, in CF '06. New York, NY, USA: Association for Computing Machinery, May 2006, pp. 71–82. doi: 10.1145/1128022.1128034.

[17]    S. Bajpayee, S. Sen, P. Dey, and I. Mukherjee, "A Quantum Public Key Cryptographic Scheme Using Entangled States and Grover Operator," in *Security and Privacy*, S. Mesnager, P. Stănică, and S. K. Debnath, Eds., Cham: Springer Nature Switzerland, 2025, pp. 171–181. doi: 10.1007/978-3-031-90587-2_12.

[18]    A. Yin, K. He, and P. Fan, "Quantum dialogue protocol based on Grover's search algorithms," *Mod. Phys. Lett. A*, vol. 34, no. 21, p. 1950169, Jul. 2019, doi: 10.1142/S0217732319501694.

[19]	L.-Y. Hsu, "Quantum secret-sharing protocol based on Grover's algorithm," *Phys. Rev. A*, vol. 68, no. 2, p. 022306, Aug. 2003, doi: 10.1103/PhysRevA.68.022306.

[20]	Z. Yu, "The Improved Quantum Secret Sharing Protocol Based on Grover Algorithm," *J. Phys.: Conf. Ser.*, vol. 2209, no. 1, p. 012031, Feb. 2022, doi: 10.1088/1742-6596/2209/1/012031.

[21]	S. Goutelle et al., "The Hill equation: a review of its capabilities in pharmacological modelling," *Fundamental & Clinical Pharmacology*, vol. 22, no. 6, pp. 633–648, 2008, doi: 10.1111/j.1472-8206.2008.00633.x.

[22]	X. Zhang and P. L. Rosin, "Superellipse fitting to partial data," *Pattern Recognition*, vol. 36, no. 3, pp. 743–752, Mar. 2003, doi: 10.1016/S0031-3203(02)00088-2.

[23]	D. L. Mohr, W. J. Wilson, and R. J. Freund, *Statistical Methods*. London, United Kingdom ; San Diego, CA: Academic Press, 2022.

[24]	U. Alon, *An Introduction to Systems Biology: Design Principles of Biological Circuits*. Accessed: Dec. 09, 2022. [Online]. Available: https://www.routledge.com/An-Introduction-to-Systems-Biology-Design-Principles-of-Biological-Circuits/Alon/p/book/9781439837177

[25]	R. Gesztelyi, J. Zsuga, A. Kemeny-Beke, B. Varga, B. Juhasz, and A. Tosaki, "The Hill equation and the origin of quantitative pharmacology," *Arch. Hist. Exact Sci.*, vol. 66, no. 4, pp. 427–438, Jul. 2012, doi: 10.1007/s00407-012-0098-5.

[26]	F. Ehsan Elahi and A. Hasan, "A method for estimating Hill function-based dynamic models of gene regulatory networks," *Royal Society Open Science*, vol. 5, no. 2, p. 171226, Feb. 2018, doi: 10.1098/rsos.171226.

[27]	H. Tonchev and P. Danev, "Robustness of different modifications of Grover's algorithm based on generalized Householder reflections with different phases," *Results in Physics*, vol. 59, p. 107595, Apr. 2024, doi: 10.1016/j.rinp.2024.107595.

[28]	H. Tonchev and P. Danev, "Studying the robustness of quantum random walk search on Hypercube against phase errors in the traversing coin by semi-empirical methods," in *Proceedings of 11th International Conference of the Balkan Physical Union — PoS(BPU11)*, vol. 427, SISSA Medialab, 2023, p. 175. doi: 10.22323/1.427.0175.

[29]	A. Kumar Das and R. Singh, "Theoretical derivation of super-ellipse model as approximation of physics-based implicit model for solar PV," *Solar Energy*, vol. 274, p. 112551, May 2024, doi: 10.1016/j.solener.2024.112551.

[30]	Y. Park, S. Jang, J. Lee, Y. Lee, and G. Kim, "An ellipsoidal model for generating realistic 3D facial textures," *International Journal of Computer Applications in Technology*, vol. 46, no. 1, pp. 36–44, Jan. 2013, doi: 10.1504/IJCAT.2013.051386.

[31]	S. Suthaharan, L. Sunkara, and S. Keshapagu, "Lame' curve-based signature discovery learning technique for network traffic classification," in *2013 IEEE International Conference on Intelligence and Security Informatics*, Jun. 2013, pp. 321–326. doi: 10.1109/ISI.2013.6578851.

[32]	G. Tian, Q. Yuan, T. Hu, and Y. Shi, "Auto-Generation System Based on Fractal Geometry for Batik Pattern Design," *Applied Sciences*, vol. 9, no. 11, Art. no. 11, Jan. 2019, doi: 10.3390/app9112383.

[33] J. Gielis, "A generic geometric transformation that unifies a wide range of natural and abstract shapes," *American Journal of Botany*, vol. 90, no. 3, pp. 333–338, 2003, doi: 10.3732/ajb.90.3.333.

[34] G. L. Long, Y. S. Li, W. L. Zhang, and L. Niu, "Phase matching in quantum searching," *Physics Letters A*, vol. 262, no. 1, pp. 27–34, Oct. 1999, doi: 10.1016/S0375-9601(99)00631-3.